\documentclass[journal=apchd5,manuscript=article]{achemso}
\setkeys{acs}{usetitle = true}
\usepackage[version=3]{mhchem}
\usepackage{graphicx}
\graphicspath{{./}{./pics/}}
\usepackage{float}
\usepackage{color}
\usepackage{bm}
\usepackage{amsmath}

\newcounter{Fig}

\newcommand{\be}{\begin{equation}}
\newcommand{\ee}{\end{equation}}

\author{Wei Liu}
\email{wei.liu.pku@gmail.com}
\affiliation{College of Optoelectronic Science and Engineering, National University of Defense
Technology, Changsha, Hunan 410073, China}
\alsoaffiliation{Nonlinear Physics Centre, Research School of Physics and Engineering, Australian National
University, Canberra, ACT 0200, Australia}
\author{Andrey E. Miroshnichenko}
\affiliation{Nonlinear Physics Centre, Research School of Physics and Engineering, Australian National
University, Canberra, ACT 0200, Australia}
\author{Yuri S. Kivshar}
\affiliation{Nonlinear Physics Centre, Research School of Physics and Engineering, Australian National
University, Canberra, ACT 0200, Australia}

\title{Q-factor enhancement in all-dielectric anisotropic nanoresonators}
\SectionNumbersOff
\begin{document}

%===============================================================

%=================================================================================
\newpage
\begin{abstract}

\textbf{It is proposed and demonstrated that Q-factor of optical resonators can be significantly enhanced by introducing an extra anisotropic cladding. We study the optical resonances of all-dielectric core-shell nanoresonators and reveal that radially anisotropic claddings can be employed to squeeze more energy into the core area, leading to stronger light confinement and thus significant Q-factor enhancement. We further show that the required homogenous claddings of unusual anisotropy parameters can be realized through all-dielectric multi-layered isotropic structures, which offers realistic extra flexibilities of resonance manipulations for optical resonators.}
\end{abstract}

\newpage

Optical resonators serve almost ubiquitously as an indispensable platform for efficient light-matter interactions and constitute the cornerstone of many related fields~\cite{Vahala2004_book_microcavity}. Different kinds of optical resonators  correspond to different light confining mechanisms, which can be roughly categorized as: total internal reflection for whispering gallery resonators~\cite{Lam1992_JOSAB_Explicit,Yariv2006_book_photonics,Vahala2004_book_microcavity}, photonic bandgap for photonic crystal cavities~\cite{akahane2003_nature_high,Joannopoulos2008_book}, Anderson localization for random resonators~\cite{abrahams2010_book_anderson,segev2013anderson}, electromagnetic surface waves for plasmonic and graphene resonators~\cite{Boardman1982_book,Maier2007,min2009_nature_high,koppens2011_NL_graphene,Liu2014_arXiv_Geometric,li2015_OL_tunable}, \textit{etc}. For optical resonator based fundamental research and applications, such as cavity-enhanced spectroscopy and sensing~\cite{Gagliardi2014_book_cavity,zhu2010_NP_chip}, cavity optomechanics~\cite{aspelmeyer2014_RMP_cavity}, a nonlinear and quantum optics~\cite{slusher1985_PRL_observation,haroche2006_book_exploring,srinivasan2007_nature_linear}, high-Q resonators are required to achieve a significant field confinement and low scattering loss rate, thus providing  an efficient platform for strong light-matter interactions.

To support high-Q resonances, the aforementioned mechanisms have different specific problems: photonic bandgap and random resonators are limited by fabrication and can not function beyond the diffraction limit; plasmonic and graphene resonators can be scaled down to the subwavelength spectral regime but, unfortunately, their performance is restricted by intrinsic losses of materials employed. Though extremely high-Q whispering gallery resonator is almost free from intrinsic loss and can be easily fabricated, it relies on the excitation of the high-order modes (higher-order mode corresponds to larger transverse light momentum, which leads to lower radiation losses and higher Q-factor~\cite{Lam1992_JOSAB_Explicit,Johnson1993_JOSAA_Theory,Liu2014_arXiv_Geometric}) and, thus, is inevitably compromised by large footprint of the resonator and also large mode volume. It was found that the principle of total internal reflection (TIR) can be extended to the interface of isotropic and anisotropic materials~\cite{chew1995_IEEE_waves,Fan2011_PRL_transverse} (or the so-called \textit{relaxed total internal refection}~\cite{Jahani2014_Optica_transparent,Jahani2015_JOSAA_photonic}, RTIR) and has been demonstrated that material anisotropy can be employed to accelerate the decay rate of evanescent waves, resulting in stronger light confinement capability in waveguiding systems~\cite{Fan2011_PRL_transverse,Jahani2014_Optica_transparent,Jahani2015_JOSAA_photonic}.

%-------------------------------------------------------------------------------
\begin{figure}
\centerline{\includegraphics[width=12cm]{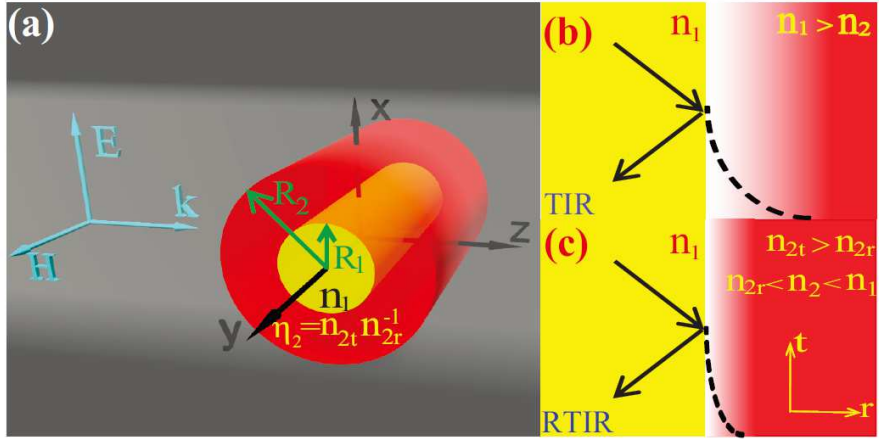}}\caption{ (a) Schematic of the core-shell cylindrical resonator consisting of an isotropic core (refractive index $n_1$ and radius $R_1$) and a radially anisotropic cladding layer (radial index $n_{2r}$, transverse index  $n_{2t}$ and radius $R_2$). The anisotropy parameter is defined as $\eta_2=n_{2t}/n_{2r}$ and the incident plane wave to excite the TM resonances is polarized along $x$ direction. (b) Conventional TIR:  the evanescent wave (dashed curve) resides within the lower-index ($n_2$) isotropic medium and decays exponentially away from the boundary. (c) RTIR: the evanescent wave decays much faster away from the boundary, which is due to the anisotropy of the medium involved.}
\label{fig1}
\end{figure}
%-------------------------------------------------------------------------------

In this paper, we study the resonances of all-dielectric core-shell cylindrical nanoresonators of nanowires from new perspectives of the total internal reflection at the interface of anisotropic media. To take advantage of the faster evanescent wave decay induced by the RTIR condition, we study a resonator that consists of an isotropic core and radially anisotropic cladding layer (the refractive index of the core is larger and smaller than the radial and transverse refractive index of the cladding layer, respectively). We show that such an anisotropic cladding can squeeze more energy of the mode into the core area, which results in  stronger field confinement and thus larger Q-factors for resonances of various orders of the resonator. We further demonstrate that the cladding layer of naturally inaccessible anisotropy can be substituted by realistic multi-layered isotropic metamaterial structures without compromising the property of significant Q-factor enhancement. The principle we have revealed is general and renders new possibilities for manipulations of resonances in terms of Q-factor enhancement, which can play a critical role in many related research directions and applications, such as cavity-enhanced spectroscopy and sensing, cavity optomechanics, resonant quantum and nonlinear optics, lasing and imaging, \textit{etc}.

\section{Results and Discussions}

We begin with one of the most fundamental structures of two-dimensional (2D) nonmagnetic cylindrical resonator as is shown in  Figure~\ref{fig1}(a): the core layer (of radius $R_1$) is isotropic and the refractive index is $n_1$; the cladding layer (of radius $R_2$) is radially anisotropic on the $x-z$ plane with radial index of $n_{2r}$ and transverse (along the azimuthal direction) index of $n_{2t}$; the anisotropy parameter is defined as $\eta_2=n_{2t}/n_{2r}$. The resonant modes of 2D cylindrical resonators can be classified into two sets: (i) the transverse-magnetic (TM) modes with no magnetic field in the $z$ direction of propagation, and (ii) the transverse-electric (TE) modes with no electric fields along the $z$ direction~\cite{Kerker1969_book,Liu2013_OL2621,Liu2015_OL_invisible}. Since the TE resonances {\em are not affected by the radial anisotropy} of the shell layer, here we study only the TM resonances with electric fields in the $x-z$ plane. As a result, the incident plane wave is fixed to be polarized along the $x$ direction (in terms of the electric field). According to the proposed principle of RTIR~\cite{Jahani2014_Optica_transparent,Jahani2015_JOSAA_photonic}, the radial anisotropy of the cladding layer can reduce the skin depth of the evanescent waves, as is shown schematically  in Figure~\ref{fig1}(b) and (c): Figure~\ref{fig1}(b) shows the conventional TIR, where an evanescent wave (dashed curve) resides within the lower index media, and it decays exponentially away from the boundary;  Figure~\ref{fig1}(c) shows the RTIR geometry, where the evanescent waves decays faster, which is induced by the anisotropic layer employed.

%-------------------------------------------------------------------------------
\begin{figure}
\centerline{\includegraphics[width=12cm]{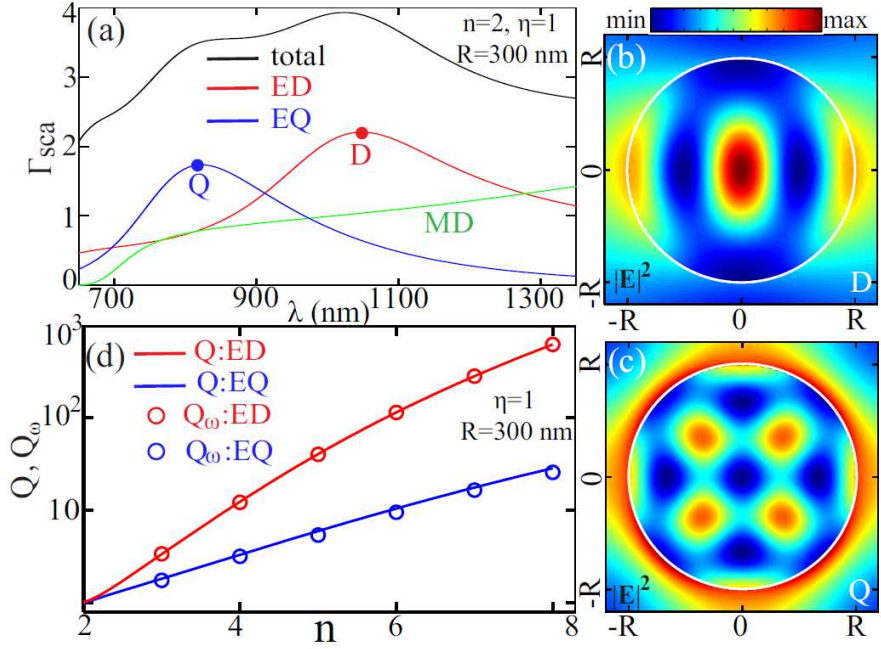}}\caption{(a) Scattering efficiency spectra for a homogeneous ($\eta=1$) and isotropic cylindrical resonator ($n=3.5$ and $R=300$~nm). Both the total scattering efficiency spectrum (black curve) and those contributions from ED (red curve), MD (green curve) and EQ (blue curve) are shown. Two points D ($\lambda_D=1046$~nm) and Q ($\lambda_Q=821$~nm), which corresponds to the resonant positions of ED and EQ respectively, are marked in (a) and the corresponding near fields (partial electric field intensity: $|\textbf{E}|^2$) are shown in (b) and (c). (d) The refractive index dependence of the Q-factors for the ED and EQ supported. The curves corresponds to the Q-factors obtained through Eq.(\ref{Quality}) while circles correspond to Q-factors calculated based on FWHM of the scattering spectra.}
\label{fig2}
\end{figure}
%-------------------------------------------------------------------------------

As a first step, we study the scattering properties of the structure  shown in Figure~\ref{fig1}(a).  This problem can be solved analytically~\cite{Kerker1969_book,chen2012_PRA_anomalous,Chen2013_OE_Tunablity}, and the scattering efficiency (scattering cross-section divided by the geometrical cross-section of the structure) can be expressed as:

%--------------------------------------------------------------
\begin{equation}
\label{Q_ext}
%Q_{\rm sca} = {2\over {kR}}[|a_0|^2 + 2\sum\nolimits_{m = 1}^\infty|a_m|^2],
\Gamma_{\rm sca} = {2\over {kR}}\sum\nolimits_{m = -\infty}^\infty|a_m|^2,
\end{equation}
%-------------------------------------------------------------
where $k$ is the angular wave number in the background material (vacuum in our work); $R$ is the radius of the outmost layer; $a_0$ and $a_m$ ($a_m=a_{-m}$) are the scattering coefficients, which depend on the anisotropy parameter (see the  Methods section for more details). To be more specific, $a_0$ corresponds to the magnetic dipole (MD), which has all the electric fields along the transverse direction on the $x-z$ plane~\cite{Liu2013_OL2621}.  As the result, the MD resonance is only dependent on the transverse (azimuthal) refractive index of the cylindrical resonator, and consequently it is not affected by the radial anisotropy.  While $a_m$ ($m\neq0$) corresponds to the electric resonance of the m$-th$ order [\textit{e.g.}, $a_1$ and $a_2$ correspond to the electric dipole (ED) and electric quadrupole (EQ), respectively].  In Figure~\ref{fig2}(a) we show the scattering efficiency spectra for a homogenous and isotropic cylinder (radius $R=300$~nm and refractive index $n=2$) in terms of the total scattering together with the contributions from the first three dominant resonances: MD (green curve), ED and EQ. As MD is not affected by the radial anisotropy, throughout this work we focus  only on the ED and EQ resonances which are both anisotropy sensitive. The indicated points D and Q in Figure~\ref{fig2}(a) correspond to the resonant positions of ED and EQ respectively: $\lambda_D=1046$~nm and $\lambda_Q=821$~nm. The corresponding near-field distributions $|\textbf{E}|^2$ at those two points are shown in Figure~\ref{fig2}(b) and (c). We show here only the partial fields of the ED and EQ resonances (fields associated with other multipoles are neglected), which as a result exhibit typical pure ED and EQ characteristics respectively.

%-------------------------------------------------------------------------------
\begin{figure*}
\centerline{\includegraphics[width=16cm]{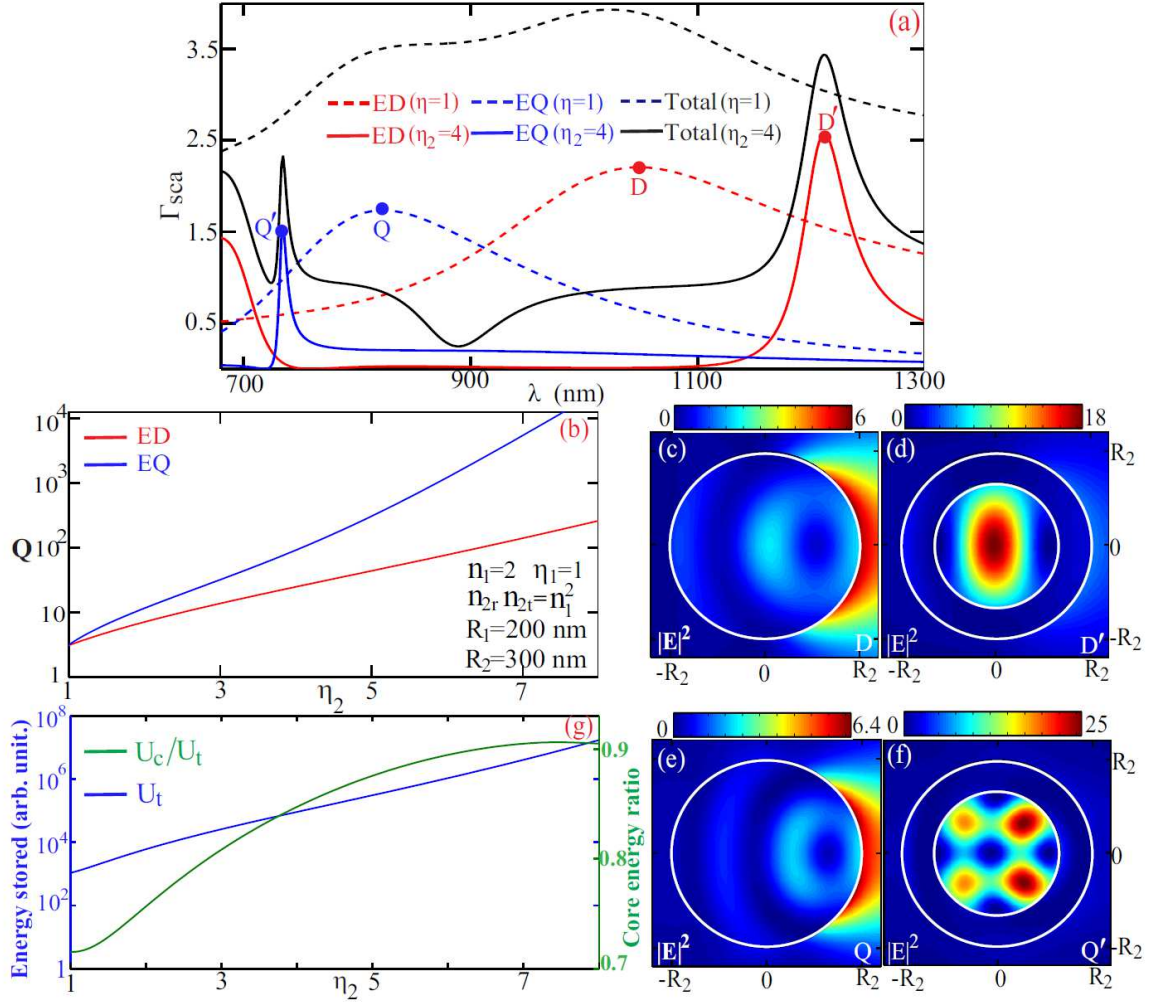}}\caption{(a) Scattering efficiency spectra for both ED (red curve) and EQ (blue curve) supported by the isotropic core-anisotpric shell cylindrical resonator of inner radius $R_1=200$~nm ($n_1=2$)  and outer radius $R_2=300$~nm ($\eta_2=4$, $n_{2t}=4$). For comparison, the ED and EQ scattering efficiency spectra for the homogenous and isotropic ($n=2$) resonator of radius $300$~nm are also shown as dashed curves.  The total scattering efficiency spectrum are shown as black curves (dashed for the isotropic case and solid for the anisotropic case). Four resonant points for both cases $D,~D'$ and $Q,~Q'$ ($\lambda_{D}=1046$~nm, $\lambda_{D'}=1213$~nm, $\lambda_{Q}=821$~nm, $\lambda_{Q'}=734$~nm) are indicated and the corresponding total near-field distributions ($|\textbf{E}|^2$) are shown in (c)-(f). (b) The dependence of the Q-factor for both ED and EQ on anisotropy parameter. The indexes of the cladding layer is constrained by $n_{2r}n_{2t}=n_1^2=4$ and thus the anisotropy parameter is $\eta_2=(n_{2t}/n_1)^2$. (g) The dependence of $U_t$ and $U_c/U_t$ for ED on the anisotropy parameter.}
\label{fig3}
\end{figure*}
%-------------------------------------------------------------------------------

Then, we study the Q-factors of the ED and EQ resonances of the homogeneous cylindrical nanowire.  If we use $
\omega _{\rm{m}} {\rm{ = }}\omega _{\rm{m}}'{\rm{ + }}i\omega _{\rm{m}}''$ to denote the complex resonant angular-frequency of the m$-th$ order resonance, which corresponds to the singular point of the scattering matrix of the resonator, then the corresponding Q-factor can be expressed as~\cite{Kerker1969_book,Yariv2006_book_photonics}:
%--------------------------------------------------------------
\begin{equation}
\label{Quality}
Q_{\rm m} =\frac{\omega _{\rm{m}}'}{2\omega _{\rm{m}}''}.
\end{equation}
%-------------------------------------------------------------
It is worth noticing that for a homogenous cylindrical resonator, the scattering matrix is dependent only on the normalized radius $\rho =kR$ and refractive index $n$.  That is to say, for fixed mode order $m$, $\omega _{\rm{m}}$ is proportional to the resonator radius.  As a result, according to Eq.(\ref{Quality}), $Q_{\rm m}$ is $R$-independent and depends only on the refractive index $n$.  The dependence of the Q-factors of ED and EQ on refractive index $n$ is shown in Figure~\ref{fig2}(d).  As is expected, with increasing $n$ the Q-factors of both ED and EQ increases monotonically, since larger momentum mismatch between the resonator and the background leads to stronger energy confinement capability~\cite{Kerker1969_book,Yariv2006_book_photonics,yang2012_NP_experimental,Liu2014_arXiv_Geometric}.  We note that besides the definition shown in Eq.(\ref{Quality}) [the results have been shown by curves in Figure~\ref{fig2}(d)], Q-factors can be also obtained via calculating the full width at half maximum (FWHM) of the scattering spectrum~\cite{Yariv2006_book_photonics}: $
{\rm{Q}}_\omega = \omega _0 /\Delta \omega _{1/2}$, where $\omega _0$ denotes the central resonant scattering frequency [such as the points indicated in Figure~\ref{fig2}(a)]; and  $\omega _{1/2}$ is width of the scattering spectrum curve measured between the points where the magnitudes are half of the maximum amplitude at the central resonant position. The calculated results through scattering spectrum are shown also in Figure~\ref{fig2}(d) by circles and it is obvious they agree well with those obtained through Eq.(\ref{Quality}).  Throughout this paper, we focus on the ED and EQ resonances of the lowest frequency [besides the resonant peaks shown in Figure~\ref{fig2}(a), there are other scattering peaks for ED and EQ at higher frequencies~\cite{Kerker1969_book,Liu2015_OE_Ultra}] and calculate the Q-factor through Eq.(\ref{Quality}). Nevertheless, it should be reminded that for ED and EQ of higher frequencies and for other ways to define and/or calculate the Q factors (or non-Lorentzian asymmetric line-widths of the scattering spectrum~\cite{Ruan2013_JPCC7324,Tribelsky2015_arXiv}), the conclusions drawn in our work are still valid.

\begin{figure*}
\centerline{\includegraphics[width=16cm]{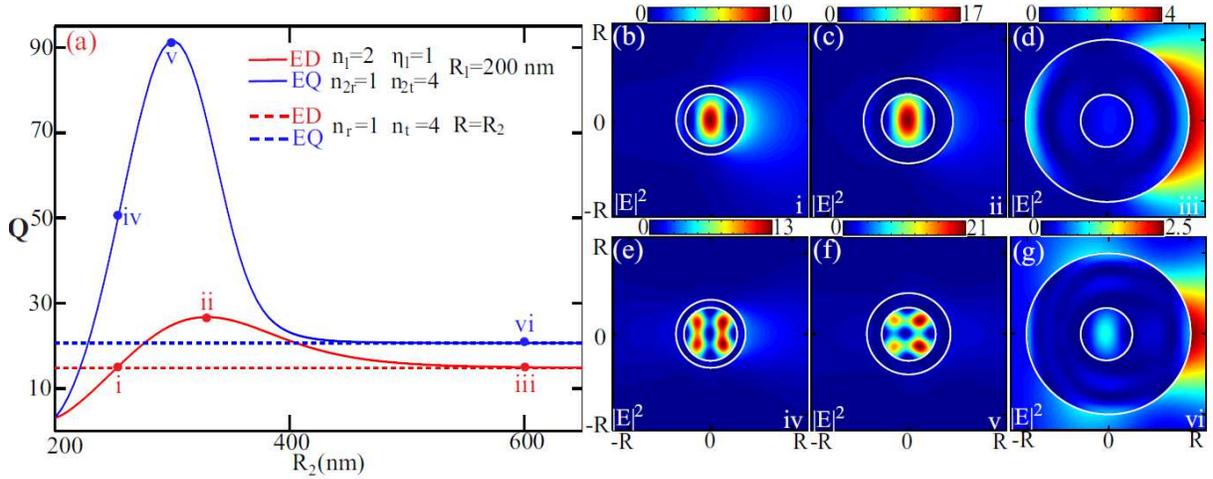}}\caption{(a) Dependence of Q-factor (solid red curve for ED and solid blue curve for EQ) on the shell cladding radius $R_2$ with the following fixed parameters: $R_1=200$~nm, $n_1=2$, $\eta_2=4$ and $n_{2t}=4$.  The dashed curves corresponds to the Q-factors (which is radius independent) for both ED (red) and EQ (blue) of the homogeneously anisotropic cylindrical resonator of $\eta=4$ and $n_{t}=4$. Three points have been indicated for the core-shell configuration [ED resonance: $R_2(\rm\romannumeral1)=250$~nm, $R_2({\rm\romannumeral2})=330$~nm, $R_2(\rm\romannumeral3)=600$~nm; EQ resonance:$R_2(\rm\romannumeral4)=250$~nm, $R_2(\rm\romannumeral5)=300$~nm, $R_2(\rm\romannumeral6)=600$~nm] and the total near-field distributions in terms of electric field intensity $|\textbf{E}|^2$ at those points are shown in (b)-(g) where $R=600$~nm.}.
\label{fig4}
\end{figure*}

\begin{figure*}
\centerline{\includegraphics[width=16cm]{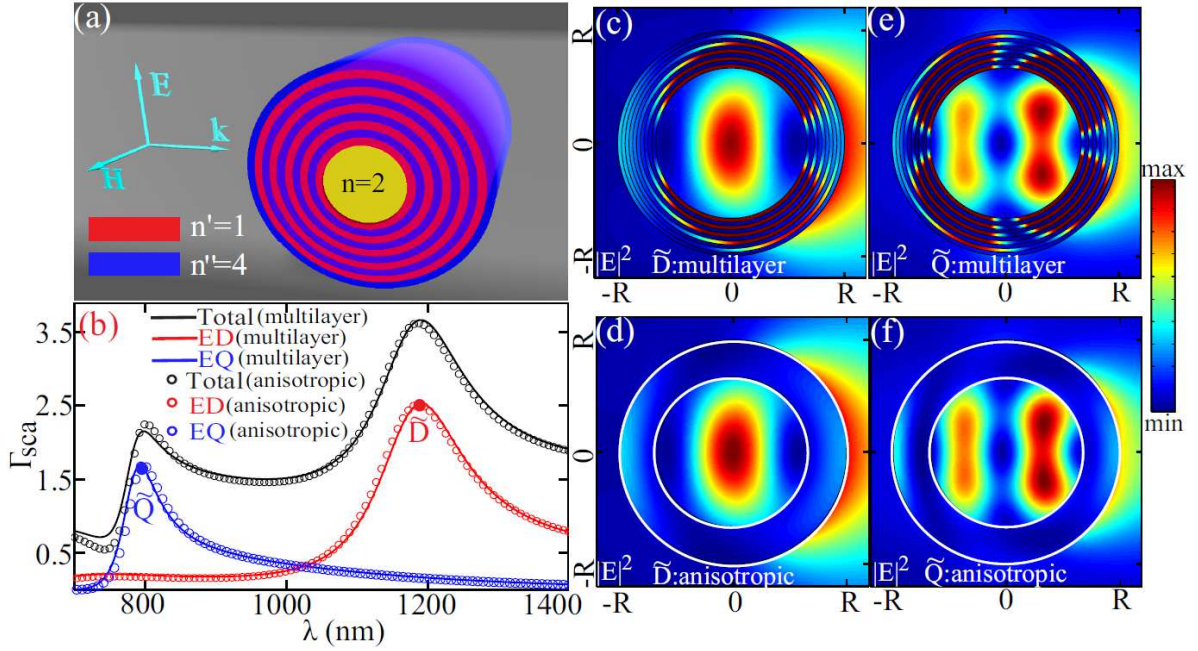}}\caption{(a) Schematic of the  cylindical resonator consisting of an isotropic core (refractive index $n_1=2$ and radius $R_1=200$~nm) and a multi-layered cladding made of alternate $n=1$ and $n=4$ isotropic layers and the width of both layer is $10$~nm. There are five layers of each medium, which leads to $f=0.5$ and $R_2=300$~nm. (b) Scattering efficiency spectra of the cylindrical resonator shown in (a) for both ED (red solid curves) and EQ (red solid curves).  The circles corresponds to the scattering efficiency spectra of a two-layered cylindrical resonator with the same core as that shown in (a) and a homogenous anisotropic cladding layer of $n_{\rm {2r}}=1.37$, $n_{\rm {2t}}=2.92$ and $R_2=300$~nm: both the spectra of ED (red circles) and EQ (blue circles) are shown. The total scattering efficiency spectrum is also shown by black curve (multilayer) and black circles (anisotropic cladding). Two points $\tilde{D}$ ($\lambda_{\tilde{D}}=1191$~nm) and $\tilde{Q}$ ($\lambda_{\tilde{Q}}=791$~nm), which correspond to the resonant positions of ED and EQ respectively, are marked in (b) and the corresponding total near-field distributions ($|\textbf{E}|^2$) are shown in (c) and (e) for the multi-layered isotropic cylindrical resonator, and in Figure~\ref{fig5}(d) and (f) for thetwo-layered core-shell  resonator with a homogenous anisotropic cladding.}
\label{fig5}
\end{figure*}

As a next step, we switch to the cylindrical resonator with an anisotropic cladding and demonstrate how the RTIR mechanism can be employed to enhance the Q-factor and energy confinement ability of the resonances. Firstly we study the core-shell resonator with an isotropic core ($R_1=200$~nm) and an anisotropic cladding. For better comparison with the homogenous resonator investigated above, we set the core index as $n_1=2$, cladding layer radius as $R_2=300$~nm and the indexes of the cladding satisfy: $n_{2r}n_{2t}=n_1^2$. As a result, the anisotropy parameter can be expressed as: $\eta_2=(n_{2t}/n_1)^2$.  The scattering spectra of $\eta_2=(n_{2t}/n_1)^2=4$ is shown in Figure~\ref{fig3}(a), where only the contributions from ED and EQ are shown [for clearer comparison, results of homogenous case with the same outmost layer radius $R=300$~nm shown already in Figure~\ref{fig2}(a) are re-plotted here by dashed curves]. As is expected from the RTIR principle, the anisotropic cladding improves the energy confinement and thus can enhance the Q-factor for both ED and EQ, which can be justified by the presence of sharper and narrower scattering curves with anisotropy materials incorporated~\cite{Kerker1969_book,Yariv2006_book_photonics}. To quantitatively characterize the Q-factor enhancement, in Figure~\ref{fig3}(b) we show the dependence of Q-factors of both resonances on the anisotropy parameter, which further verifies the anisotropy-origin of such enhancement. In principle, higher Q-factor indicates better energy confinement capability of the resonator. To confirm this directly,  in Figure~\ref{fig3}(a), the positions of the ED and EQ resonances for both isotropic and anisotropic cases are indicated by $D,~D'$ and $Q,~Q'$ ($\lambda_{D}=1046$~nm, $\lambda_{D'}=1213$~nm, $\lambda_{Q}=821$~nm, $\lambda_{Q'}=734$~nm) and the corresponding total electric field (combining the contributions of all multipoles) intensity ($|\textbf{E}|^2$) are shown in  Figure~\ref{fig3}(c)-(f).  Compared to the isotropic case [Figure~\ref{fig3}(c) and (e)], when anisotropy is introduced [Figure~\ref{fig3}(d) and (f)] more energy of the mode is squeezed into the core-layer, indicating better energy confinement and as a result leading to more significant field enhancement [we note that to make the field distributions clearer, different scales of the color-bars for the fields at different points are employed in Figure~\ref{fig3}(c)-(f)]. It worth mentioning that compared to the symmetric field distributions shown in Figure~\ref{fig2}, those shown in \ref{fig3} are asymmetric as here we show the total fields contributed by all the multipoles.  This is also the case for \ref{fig4}-\ref{fig5}.

To further quantify the capability of better energy confinement and more significant field enhancement, we denote the energy stored inside the core and shell cladding as $U_c$ and $U_s$, which can be defined respectively as:
\begin{equation}
\label{Energy}
U_{\rm c}  = \mathop{\int\!\!\!\int\!\!\!\int}\limits_{\rm core}{n_1^2 |E(\textbf{r})|^2 d^3\textbf{r}},~~~
U_{\rm s}  = \mathop{\int\!\!\!\int\!\!\!\int}\limits_{\rm shell}{(n_{2r}^2 |E_r (\textbf{r})|^2  + n_{2t}^2 |E_t (\textbf{r})|^2 )d^3\textbf{r}},
\end{equation}
where $E_r (\textbf{r})$ and $E_t (\textbf{r})$ correspond to electric fields along the radial and transverse direction respectively, which can be calculated analytically~\cite{Kerker1969_book,chen2012_PRA_anomalous,kim2015_SR_invisible};  the total stored energy inside the resonator is $U_{\rm t}=U_{\rm c} +U_{\rm s}$; and the core energy ratio is $U_{\rm c}/U_{\rm t}$.  In Figure~\ref{fig3}(g) we show the dependence of $U_{\rm t}$ and $U_{\rm c}/U_{\rm t}$ on the anisotropy parameter $\eta_2$ for the ED resonance. As is clearly shown, with larger $\eta_2$  both the total stored energy and the core energy ratio would increase, which proves convincingly tighter energy confinement and larger field enhancement inside the resonator that accompany the efficient Q-factor enhancement shown in Figure~\ref{fig3}(a).

In the discussions above, we fixed the overall resonator radius to investigate the Q-factor dependence on anisotropy parameter. Now we fix the anisotropy parameter $\eta_2=4$ (with $R_1=200$~nm, $n_1=2$ and $n_{2t}=4$) and study the relationship between Q-factor and the shell cladding radius $R_2$. The results are summarized in Figure~\ref{fig4}(a) as solid curves for both ED and EQ resonances. It is clear that the Q-factor does not increase monotonically with increasing cladding radius, and there is actually an optimal radius of the shell to achieve the highest Q-factor for each resonance. To reveal the mechanism behind, we select three points for both ED [$R_2(\rm\romannumeral1)=250$~nm, $R_2(\rm\romannumeral2)=330$~nm (optimal radius for EQ), $R_2(\rm\romannumeral3)=600$~nm] and EQ [$R_2(\rm\romannumeral4)=250$~nm, $R_2(\rm\romannumeral5)=300$~nm (optimal radius for ED), $R_2(\rm\romannumeral6)=600$~nm], and show the corresponding total near-field distributions ($|\textbf{E}|^2$) in Figure~\ref{fig4}(b)-(g). It is shown that before the optimal radius, for both resonances more energy is squeezed into the core region with increasing cladding radius, which leads to both Q-factor and near-field enhancement. In a sharp contrast, after the optimal radius, with increasing $R_2$ more and more energy will leak out from the core, reducing both the Q-factor and the field intensity within the resonator. This is due to the fact that when the cladding layer is becoming sufficiently large, resonances can be formed within the cladding layer, with comparably negligible energy distributed within the core layer [see Figure~\ref{fig4}(d) and (g)]. To confirm directly this process of mode distribution transformation between the core and the shell cladding, we plot the Q-factor-radius relationship of a homogeneous anisotropic cylindrical resonator with $\eta=4$ and $n_{2t}=4$ for both ED (dashed red curve) and EQ (dashed blue curve) in Figure~\ref{fig4}(a): it is obvious that for sufficiently large $R_2$ the Q-factor would converge to that of a homogeneous anisotropic cylinder [it is worth mentioning that similar to the isotropic case studied in Figure~\ref{fig2}, the Q-factor of an anisotropic homogeneous resonator is also independent on the radius]. The results presented here could be confusing at a first glance, as it is taken for granted that increasing the anisotropic layer width would not break the condition of RTIR and thus would not result in reduced Q-factor or near-field intensity. But one has to keep in mind that RTIR (and TIR itself) is a  concept of geometric optics, and thus cannot be applied in a universal way. When the cladding layer is sufficiently large, potential distributions for photons would be changed drastically~\cite{Johnson1993_JOSAA_Theory}, resulting in more energy confinement within the cladding layer and undermining the features of Q-factor and near-field enhancement.

%-------------------------------------------------------------------------------
%-------------------------------------------------------------------------------

%-------------------------------------------------------------------------------
%-------------------------------------------------------------------------------

Up to now, we have demonstrated the efficient Q-factor enhancement based on anisotropic materials. Unfortunately, for natural materials the anisotropy parameters employed here may not be realistic. Such a problem is not insurmountable considering the recent development in the field of artificial metamaterials, where other extreme refractive indices and anisotropy parameters can be achieved~\cite{Soukoulis2011_NP,Zheludev2012_NM,choi2011_nature_terahertz,yang2012_NP_experimental,wu2014_PRX_electrodynamical,poddubny2013_NP_hyperbolic,kim2015_SR_invisible}. Here, we employ a cladding layer consisting of multilayers of two kinds of realistic isotropic materials ($n'$ and $n''$ respectively and $n'<n''$ ) to substitute the homogeneous anisotropic layer, as shown in Figure~\ref{fig5}(a).  According to the effective medium theory~\cite{agranovich1985_SSC_notes,poddubny2013_NP_hyperbolic,kim2015_SR_invisible}, the effective indexes for the multi-layered cladding in the $x-z$ plane are:

\begin{equation}
\label{effective1}
n_{\rm {2r}} =n'n''/\sqrt {(1 - f)n'^2  + fn''^2 },~~~n_{\rm {2t}} =\sqrt {fn'^2  + (1 - f)n''^2 }, %
\end{equation}
where $f$ is the filling factor of the layer of index $n'$. When $f=0.5$ the largest anisotropy parameter can be achieved: $\eta_2={\kern 1pt} (n'^2  + n''^2 )/2n'n''$ for positive $n'$ and $n''$, and $n_{2r}n_{2t}=n'n''$. As a proof of concept, we set $n'=1$ and $n''=4$ (\textit{e.g. }for Germanium), which leads to the following parameters for $f=0.5$: $n_{\rm {2r}}=1.37$, $n_{\rm {2t}}=2.92$  and  $\eta_2=2.13$. The structure we employ is shown schematically in Figure~\ref{fig5}(a), where the cladding is made of alternate $n'$ and $n''$ layers and the width of both layer is $10$~nm [which is far smaller then the effective wavelength and then the effective medium theory Eq.(\ref{effective1}) can be applied]. There are five layers of each medium, which makes $f=0.5$ and the overall cladding layer of radius $R_2=300$~nm (for direct comparisons with the previous discussions, the core is set to be isotropic with radius $R_1=200$~nm and of $n_1=2$). The scattering efficiency spectra of such isotropic $11$-layered  cylindrical resonator can be calculated analytically~\cite{Kerker1969_book} and the results are shown in Figure~\ref{fig5}(b) by solid curves for both ED and EQ resonances. It is worth mentioning that here the periodicity of the cladding is not important as far as the layer width is far smaller than the wavelength when the effective medium theory can be applied. This distinguishes our design from those based on photonic bandgap structures~\cite{akahane2003_nature_high,Joannopoulos2008_book}. To verify the effective medium theory applied for the multi-layered cladding, we also show in Figure~\ref{fig5}(b) by circles the scattering efficiency spectra when the multi-layered cladding is replaced by a homogenous anisotropic layer [according to Eq.(\ref{effective1})] of $n_{\rm {2r}}=1.37$, $n_{\rm {2t}}=2.92$, and $n_{2r}n_{2t}=n_1^2=4$. It is obvious that both sets of results agree excellently well, indicating the same Q-factor for both cases [this has already been included in Figure~\ref{fig3}(b)].  The two indicated points $\tilde{D}$ and $\tilde{Q}$ in Figure~\ref{fig5}(a) correspond to the resonant position of ED and EQ respectively: $\lambda_{\tilde{D}}=1191$~nm and $\lambda_{\tilde{Q}}=791$~nm. The corresponding total near-field distributions ($|\textbf{E}|^2$) are shown in Figure~\ref{fig5}(c) and (e) for the multi-layered isotropic cylindrical resonator, and in Figure~\ref{fig5}(d) and (f) for the core-shell anisotropic resonator. The strong field confinement is clearly demonstrated by comparing Figure~\ref{fig5}(c-f) with Figure~\ref{fig3}(c) and (e). Although the field distributions within the cladding layer are contrastingly different (the fields are significantly enhanced within the $n=1$ layers due to the continuity condition along the radial direction, which is similar to that shown in Ref.~\cite{Oulton2008_NP_hybrid}), the fields inside the core and outside the resonator are almost the same, proving further the feasibility of applying the RTIR principle to obtain high-Q cavity resonators.

\section{Conclusion and Outlook}

In summary, based one the recently proposed principle of light localization in anisotropic media, we have studied all-dielectric cylindrical resonators with an isotropic core and anisotropic cladding, focusing on the Q-factor enhancement of the resonances supported.  Due to the RTIR effect, we have demonstrated that more  energy can be confined within the core-layer structure, which leads to simultaneous Q-factor and near-field enhancement for the resonances. We have also shown that with a fixed radial anisotropy parameter of the cladding, there exist  optimal cladding layer widths to achieve the highest Q-factors for resonances of different orders, thus succeeding in revealing the limitation of the RTIR principle.  To verify the feasibility of our approach to achieve high-Q resonators, we have shown that  naturally inaccessible anisotropy parameters of the  cladding can be realized by employing multi-layered isotropic metamaterial structures with the property of significant Q-factor enhancement preserved.

It is worth mentioning that though in this work we confine our study to the two lowest order of modes (dipole and quadruple modes) of cylindrical structures, the principle we have demonstrated is quite general. The Q-factor enhancement can certainly be achieved in spherical resonators and resonators of other shapes, for other kinds of anisotropy such as magnetic anisotropy and for modes of higher orders.  Actually the higher the mode order is, the stronger the Q-factor enhancement effect would be [see Figure~\ref{fig3}(b)]. Since more energy is confined within the core of the resonator by the RTIR effect, it is expected that for a cluster of such resonators the near-field coupling between them can significantly be reduced~\cite{Liu2012_ACSNANO,Miroshnichenko2012_NL6459}. For the proof of concept demonstration, we have employed the multi-layered positive-index isotropic cladding with modest anisotropy parameter of $\eta_2=2.125$. Other more unusual anisotropy parameters, even including those complex ones, can be made available for realistic applications relying on the artificial metamaterials~\cite{Soukoulis2011_NP,Zheludev2012_NM,choi2011_nature_terahertz,yang2012_NP_experimental,wu2014_PRX_electrodynamical,poddubny2013_NP_hyperbolic,kim2015_SR_invisible}.  We believe that our work of employing  effective anisotropic media to enhance the Q-factor opens a new dimension of freedom for  manipulations of various resonators, which can incubate many new resonator based fundamental research and applications in the fields of cavity-enhanced spectroscopy and sensing, cavity optomechanics, resonant quantum and nonlinear optics, lasing and imaging, \textit{etc}.

\section{Methods}

The seminal problem of two-dimensional (2D) plane wave scattering by cylindrical resonators (single or multilayered) consisting of isotropic materials has been studied analytically~\cite{Kerker1969_book} [\textit{e.g.}, for the scattering configuration shown in \ref{fig5}(a) with the corresponding scattering efficiency shown in \ref{fig5}(b) by solid curves]. Such analytical investigations have also been extended to 2D cylindrical resonators  made of radially anisotropic materials~\cite{chen2012_PRA_anomalous,Chen2013_OE_Tunablity} [\textit{e.g.}, for geometries shown in \ref{fig1}(a)]. It has been revealed that for both isotropic and radially-anisotropic cases, the expressions for the field expansion coefficients (both inside and outside the resonator) and  far-field scattering properties are identical [see Eq.(\ref{Q_ext})], except that in the anisotropic case the orders of some Bessel and Hankel functions need to be modified by the anisotropy parameter.

To be more specific, we take for example the simplest case of a homogeneous radially anisotropic cylinder (radius $R$ and anisotropy parameter $\eta$) in vacuum. The scattering efficiency can be expressed by Eq.(\ref{Q_ext}) where the scattering coefficients for the incident plane waves of TM polarizations can be expressed as~~\cite{Kerker1969_book,chen2012_PRA_anomalous}:

\begin{equation}
\label{method_1}
a_m  =a_{-m}={{n_tJ _{\tilde m} (n_tkR )J '_{m} (kR ) - J _{m} (kR )J '_{\tilde m} (n_tkR )} \over {n_tJ _{\tilde m} (n_tkR )H '_{m} (kR ) - H _{m} (kR )J '_{\tilde m} (n_tkR )}}~~~(m\geq0),
\end{equation}
where $\tilde m$ is the modified function order $\tilde m = m\eta$, $J$ and $H$ are the Bessel and Hankel functions of the first kind~\cite{Kerker1969_book} and the accompanying primes indicate their differentiation with respect to the entire argument.  Obviously the results will be reduced to the those of isotropic case when $\eta=1$.  It is clear that the MD mode of $m=0$ is independent of the anisotropy, which is consistent with our former argument based on field distribution.  Similar to Eq.(\ref{method_1}), all the field expansion coefficients within the resonator can be obtained in all the layers through simply modifying the function order when the radially anisotropic materials are present.  As a result the energy stored within the resonator [see \ref{Energy}] can also be calculated analytically.

\section{Acknowledgements}

We acknowledge a financial support from the National Natural Science Foundation of China (Grant number: $11404403$), the Australian Research Council and the Basic Research Scheme of College of Optoelectronic Science and Engineering, National University of Defence Technology. W. L. thanks the Nonlinear Physics Centre for a warm hospitality during his visit to Canberra.

%\bibliographystyle{achemso}
%\bibliographystyle{ol2}
%\bibliography{References_scattering}
%==========================================================References end
\providecommand{\latin}[1]{#1}
\providecommand*\mcitethebibliography{\thebibliography}
\csname @ifundefined\endcsname{endmcitethebibliography}
  {\let\endmcitethebibliography\endthebibliography}{}

\end{document}